# Numerical Analysis of Seismic Wave Amplification in Nice (France) and comparisons with experiments

## J.-F. Semblat [a], A.-M. Duval [b], P. Dangla [a]

[a] *Laboratoire Central des Ponts et Chaussées, Eng. Modelling Dept,*
*58 bd Lefebvre, 75732 Paris Cedex 15, France (semblat@lcpc.fr))*
[b] *CETE Méditerranée, Seismic risk team, 56 bd de Stalingrad, 06300 Nice, France*

The analysis of site effects is very important since the amplification of seismic motion in some specific areas can be very strong. In this paper, the site considered is located in the center of Nice on the French Riviera. Site effects are investigated considering a numerical approach (Boundary Element Method) and are compared with experimental results. The experimental results are obtained thanks to real earthquakes (weak motion) and microtremor measurements. The investigation of seismic site effects through numerical approaches is interesting because it shows the dependency of the amplification level on such parameters as wave velocity in surface soil layers, velocity contrast with deep layers, seismic wave type, incidence, damping...

In this specific area of Nice, experimental measurements obtained for weak motion lead to strong site effects. A 1D-analytical analysis of amplification does not give a satisfactory estimation of the maximum reached levels. A boundary element model is then proposed considering different wave types (SH, P, SV) as the seismic loading. The alluvial basin is successively assumed as an isotropic linear elastic medium and an isotropic linear viscoelastic solid with Zener type behaviour (standard solid). The influence of frequency and incidence is analyzed. The thickness of the surface layer, its mechanical properties, its general shape as well as the seismic wave type involved have a great influence on the maximum amplification and the frequency for which it occurs. For real earthquakes, the numerical results are in very good agreement with experimental measurements for each motion component. The boundary element method leads to amplification values very close to the actual ones and much larger than those obtained in the 1D case. Two dimensional basin effects are then very strong and are well reproduced numerically.

*Key words:* site effects, weak motion, boundary element method, numerical modelling, wave amplification, microtremors, damping.

## 1 SEISMIC WAVE PROPAGATION AND SITE EFFECTS

### 1.1 *Seismic site effects*

The amplification of seismic waves in some specific sites can be very important. Reflections and scattering of seismic waves near the surface, at layers interfaces or around topographic irregularities often strengthen the consequences of earthquakes [4,32]. It was the case for the terrific Mexico 1985 earthquake : maximum acceleration was around 0.28 g at sixty kilometers from the source and 0.2 g in some specific areas in Mexico city (lake deposits) located much further (400 kilometers). In the town itself, seismic motion was amplified up to a factor of 60 compared to the bedrock because of the soft clay deposit under the city.

Caracas 1967 earthquake has also been amplified in the center of the city because of an alluvial filling [16,38]. The resonance frequency of the basin is around 0.6 Hz in the most destroyed blocks of Caracas : the buildings that felt down in this area had 14 storeys and their resonant frequencies were very close to 0.6 Hertz. The damages are then much larger on buildings having resonant frequencies close to that leading to the stronguest site effects in the surface soil layers. It is therefore very important to avoid, in the design of structures, buildings involving resonant features close to that of the soil surface layers if it corresponds to a strong motion amplification. The local seismic response of soils must be analyzed to precisely determine the characteristics of the reference earthquake used for the design of structures.

## 1.2  *Analysis of seismic wave propagation*

Seismic wave propagation can be investigated using different kinds of experimental techniques or numerical methods. Experimental investigations are generally made at full or reduced scales :
- real earthquakes or microtremor measurements [10,14,15,16,19,34,44],
- in situ, laboratory or centrifuge experiments for soil characterization [18,22,25,26,36,39,42],

There are also various numerical methods :
- finite element method [1,5,7,20,24,27,35,37,41,45],
- boundary element method [2,6,8,13,31,38] well-adapted to the analysis of wave propagation in infinite media,
- many other types of analytical and numerical methods (spectral elements, discrete wave-number, Aki-Larner...) [4,9,17,28].

This paper investigates the local amplification of seismic waves (site effects) through a numerical model based on the boundary element method and experimental measurements (real earthquakes and microtremors). Some simple analytical results are also given to make some comparisons with experimental and numerical ones.

## 1.3  *Seismic measurements in Nice*

Nice areas having a high population density, like in many other towns, are located on alluvial soil layers. A first survey (1984) has given an estimation of the regional seismic hazard and a determination of different geological areas with potential seismic amplification. Seismicity was recorded for one year (1992) on four sites where seismic amplification is expected to be important. Experimental measurements of site effects [15] have been performed above an alluvial filling corresponding to an old valley of North-South direction axis (figure 1). Microtremor recordings stability has also been studied in both time and frequency domains. It has given the spectral variations during the day and enlightened the surprising stability with time of the spectral ratios between the horizontal and vertical compo-nents [14,15] . This ratio allows the determination of the resonant frequency of the soil surface layers.

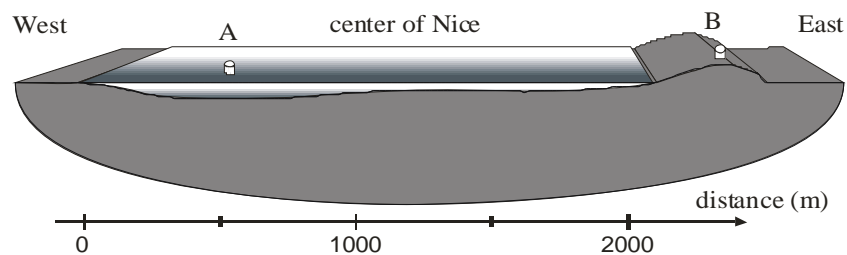

**Figure 1.** 2D geological profile in the center of Nice

The curves of figure 2 give transfer functions, between alluvial site (point A) and reference bedrock site (point B), for real earthquake measurements considering vertical, North-South and East-West (weak) motions. The experimental study [14,15] includes several seismic events. There are 7 regional earthquakes of minimum magnitude 2 among which the one leading to the largest velocity at the reference site has a magnitude of 4.6 at approximately 100 km. In the weak earthquake measurements, 4 far-field earthquakes located at more than 1000 km are also considered. Curves of figure 3 give horizontal to vertical motion ratios estimated from microtremor measurements considering both North-South and East-West horizontal motions.

Earthquake measurements, as well as microtremor analysis (figures 2 and 3), clearly indicate that the amplification of seismic motion occurs between 1 and 2 Hz at the center of the alluvial filling. The values of resonant frequency determined by both methods are very close. The amplification factor estimated from microtremor measurements is however smaller. It is not the purpose of this article to discuss the discrepancy between the amplification values obtained with both experimental methods. Some various explanations and several complete studies on microtremor techniques have already been proposed and consider both theoretical aspects as well as field observations [10,12,14,15,19,33] .

Above the thickest part of the alluvial basin, the amplification factor determined by seismic spectral ratios (site/reference) reaches a maximum value of 20 around a frequency of 2.0 Hz for vertical motion, 1.0 Hz for North-South motion and 1.3 Hz for East-West motion. For the East-West component, the amplification is not negligible for low frequencies and this is due to two near field earthquakes. Comparing the measurements performed on various sites, corresponding transfer functions show a strong dependency of resonant frequency on the thickness of the alluvial surface layers [14,15] .

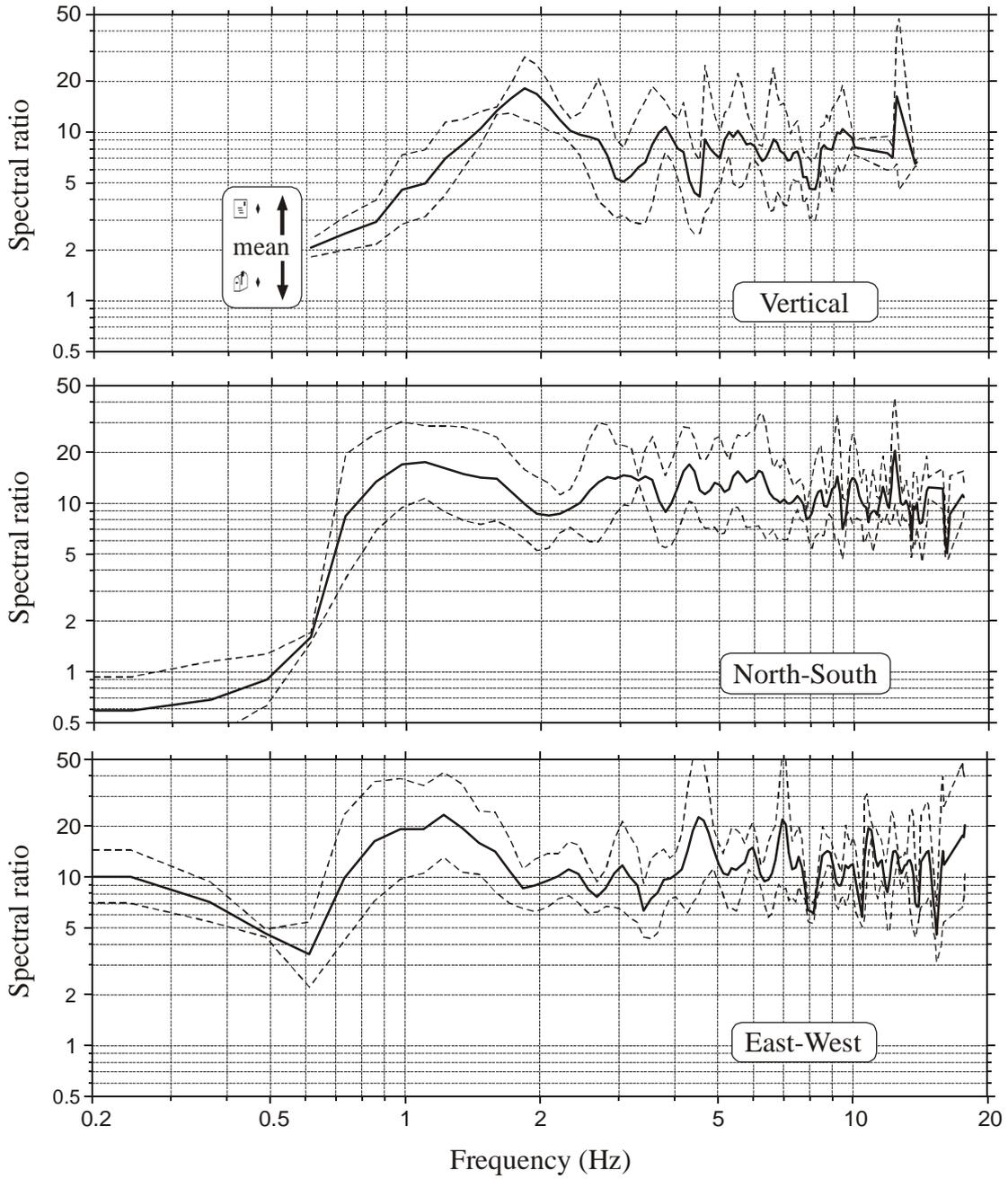

**Figure 2.** Experimental spectral ratios (site/reference) estimated
from real earthquakes measurements (weak motion).

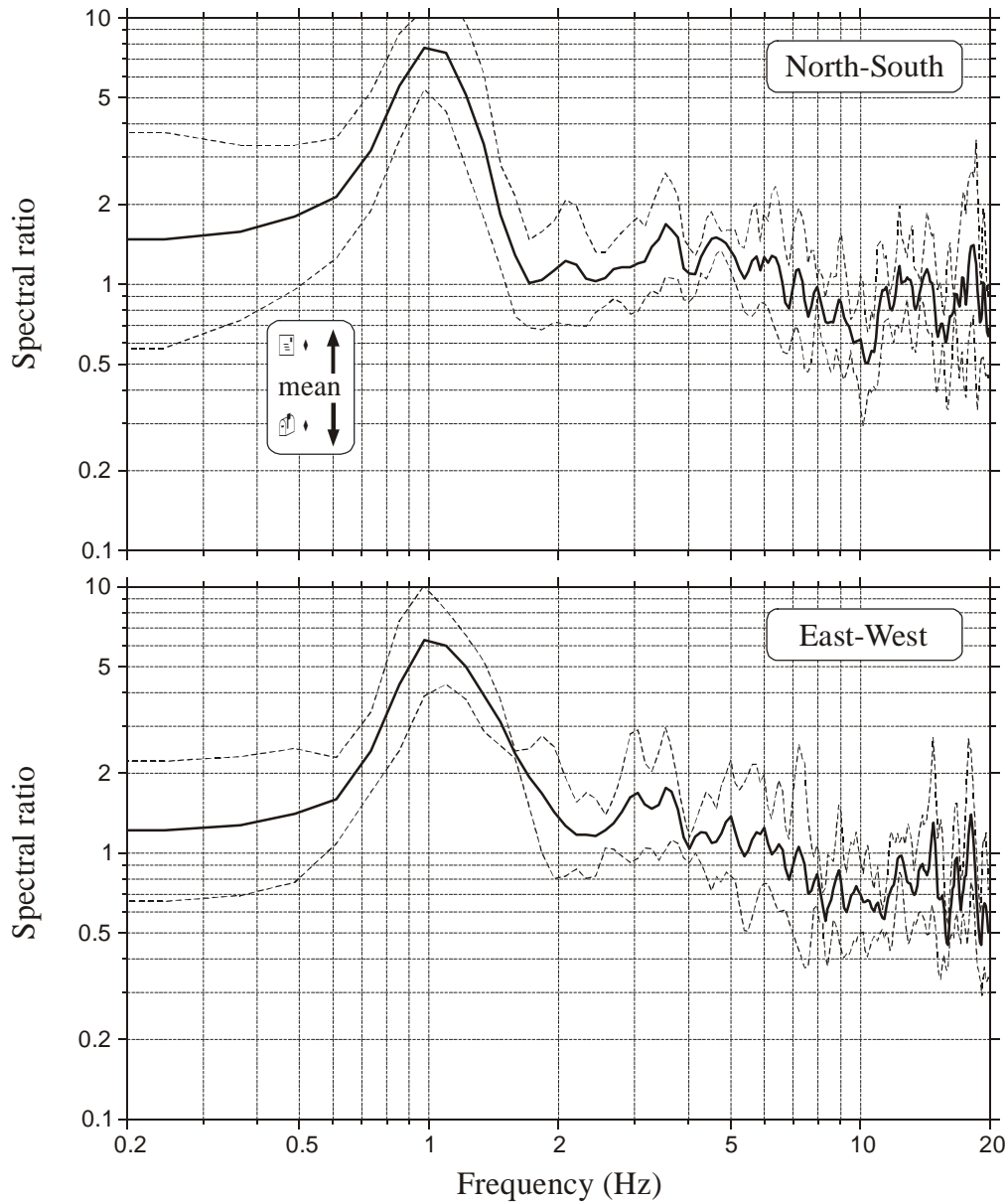

**Figure 3.** Experimental spectral ratios (H/V) estimated
from microtremors measurements.

## 1.4 *Strong motion and non linear effects*

Various experimental studies investigate the nonlinear effects in soils and strong motion responses [21,22,25,42,44] . Concerning numerical models, the finite element method [1,5], the finite difference method [22] and even the boundary element method [8] , can be used to analyze nonlinear problems. Since no strong motion recordings are available for the city of Nice, the experimental results presented in this article correspond to weak seismic motion and the numerical analysis considered in the following sections is linear. Nonlinear effects general-ly lead to a lower frequency content and can make the amplitude of the dynamic response decrease significantly. The main difficulty for the analysis of nonlinear dynamic effects is that they can strongly vary from one earthquake to another and from one site to another since the nonlinear dynamic properties of soils are influenced by many different parameters (strain amplitude, initial state, loading history...) [21,22,25, 42,44].

## 2 PRELIMINARY ANALYTICAL RESULTS

A first estimation of the amplification of the seismic motion is made considering an horizontal infinite alluvial layer of constant thickness above an elastic half-space. For this analytical and the following numerical models, an East-West geological contour is considered. For the surface layer, the velocity is chosen as a mean value of the velocities in the seven different layers of the site (above the bedrock). From different types of experimental investigations, these velocities are found to range from 170 m/s to 400 m/s [14,15]. We consequently chose the following mechanical characteristics for both media :

- alluvial layer :       $\rho_1$=2000 kg/m$^3$, $\mu_1$=180 MPa       giving    $C_1$=300 m/s,
- elastic bedrock :      $\rho_2$=2300 kg/m$^3$, $\mu_2$=4500 MPa       giving    $C_2$=1400 m/s.

where $\rho$ is the mass density, $\mu$ the shear modulus and $C$ is the shear wave velocity.

Considering a plane SH-wave with incidence $\alpha_2$ on the surface layer, the determination of the transfer function through the surface layer can be calculated analytically (figure 4). The thickness of the alluvial filling is successively : h=64 m and h'=32 m, which are the thicknesses of the two main parts of the real geological contour depicted in figure 1 (western part (thickest) and eastern part (thinnest)).

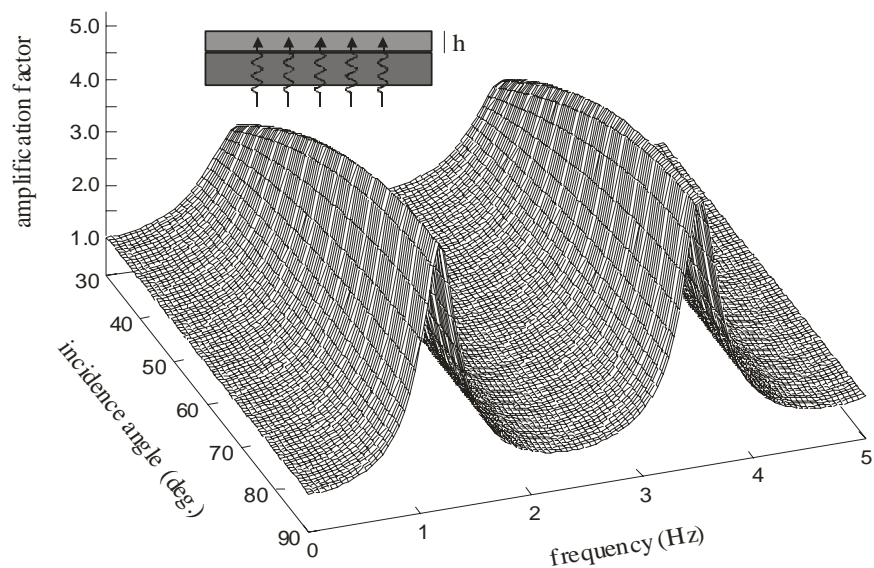

**Figure 4.** Amplification factor versus incidence and frequency for a single layer of constant thickness h=64m (analytical transfer function).

Analytical expression of the transfer function of a plane SH-wave through an elastic layer of thickness $h$ above an elastic half-space is very simple [30] :

$$\left| T_{1,2}^{*}(\omega) \right| = \left[ \cos^2 p_1 h + q^2 \sin^2 p_1 h \right]^{-1/2} \qquad (1)$$

$$\text{with } p_1 = \frac{\omega \cos\alpha_1}{C_1} \text{ et } q = \sqrt{\frac{\rho_1 \mu_1}{\rho_2 \mu_2}} \frac{\cos\alpha_1}{\cos\alpha_2}$$

where $\omega$ is the circular frequency, $\alpha_1$ and $\alpha_2$ are the angles of incidence (with respect to the vertical direction) in the layer and in the half-space (respectively).

Expression (1) directly gives the amplification factor of the seismic motion. The amplitude above the alluvial basin is already divided by the amplitude above the elastic half-space in the case there is no surface layer. From the curve of figure 4, amplification factor is around 5.5 for h=64m and normal incidence ($\alpha_2$=90°). The same value is obtained for h'=32m and the resonant frequencies are respectively :

- for the thickest part (h=64 m) :          f=1.2 Hz,
- for the thinnest part (h'=32 m) :          f'=2.3 Hz.

As shown by figure 4, the amplification factor decreases with decreasing incidence (from normal incidence of 90° down to 30°). Compared to experimental results of figure 2, the amplification factor estimated analytically is much lower (approximately four times) whereas occuring frequency values are rather good. The amplification of seismic motion in surface soft layers is not only due to the impedance ratio between elastic half-space and the alluvial deposit (considered in the analytical approach). The discrepancy between experimental and analytical results indicates that the geometrical features of the basin must have a strong influence on seismic site effects. It raises the need for a more accurate approach and in the following we will consider the boundary element method to model seismic wave propagation and amplification in the site. Other numerical or analytical methods allow the analysis of seismic wave amplification using modal approaches, finite element method, spectral element method, discrete wavenumber... [1,3,4,9,17,27,28,29,40,41,45] .

## 3 BOUNDARY ELEMENT FORMULATION

### 3.1 *The boundary element method*

The main advantage of the boundary element method is to avoid artificial truncation of the domain in the case of infinite medium. For dynamic problems, this truncation leads to artificial wave reflections giving a numerical error in the solution. The boundary element method can be divided into two main stages [6,8] :

- solution of the boundary integral equation giving displacements and stresses along the boundary of the domain,
- a posteriori computation for all points inside the domain using an integral representation formula.

The boundary element method arise from the application of Maxwell-Betti reciprocity theorem leading to the expression of the displacement field inside the domain $\Omega$ from the displacements and stresses along the boundary $\partial\Omega$ of the domain [2,6,8,13].

### 3.2 *Elastodynamics*

We consider an elastic, homogeneous and isotropic solid of volume $\Omega$ and external surface $\partial\Omega$. In this medium, the equation of motion can be written under the following form :

$$(\lambda + 2\mu)grad(div\ u) - \mu rot(rot\ u) + \rho f = \rho \ddot{u} \tag{2}$$

where $\sigma$ is the stress tensor, $u$ the displacement field and $f$ a volumic density of force.

In this article, the problem is supposed to have an harmonic dependence on time of circular frequency $\omega$. The equation of motion for a steady state $(u(x),\ \sigma(x))$ can then be written as follows :

$$(\lambda + 2\mu)grad(div\ u(x)) - \mu rot(rot\ u(x)) + \rho f(x) + \rho \omega^2 u(x) = 0 \tag{3}$$

The alluvial basin is firstly considered as a linear elastic undamped medium. Damped mechanical properties are included afterwards in the previous equation through the complex modulus of the medium.

### 3.3 *Integral representation*

For steady solutions of harmonic problems, the reciprocity theorem between two elastodynamic states $(u(x),\ \sigma(x))$ and $(u'(x),\ \sigma'(x))$ takes the following form [13] :

$$\int_{\partial\Omega} t^{(n)}(x)\ u'(x)\ ds(x) + \int_{\Omega} \rho f(x)\ u'(x)\ dv(x) =$$
$$\int_{\partial\Omega} t'^{(n)}(x)\ u(x)\ ds(x) + \int_{\Omega} \rho f'(x)\ u(x)\ dv(x) \tag{4}$$

The integral formulation is obtained through the application of the reciprocity theorem between the elastodynamic state $(u(x),\ \sigma(x))$ and the fundamental solutions of a reference problem called Green kernels. The reference problem generally corresponds to the infinite full space case in which a volumic concentrated force at point $y$ acts in direction $e$. In the harmonic case, the Green kernel of the infinite medium corresponds to a volumic force field such as :

$$\rho f'(x) = \delta(x - y)\ e \tag{5}$$

In this article, the model involves the Green functions of an infinite medium [8,13] or semi-infinite medium (in the case of SH-waves). The Green kernel is denoted $U_{ij}^{\omega}(x,y)$ and characterizes the complex displacement in direction $j$ at point $x$ due to a unit force concentrated at point $y$ along direction $i$. The corresponding stress for a surface of normal vector $n(x)$ is denoted $T_{ij}^{(n)\omega}(x,y)$. The application of the reciprocity theorem between the elastodynamic state $(u(x),\ \sigma(x))$ and that defined by the Green kernel $U_{ij}^{\omega}(x,y)$ gives the following integral representation :

$$I_{\Omega}(y)\,u_i(y) = \int_{\partial\Omega}\Big(U_{ij}^{\omega}(x,y)\,t_j^{(n)}(x) - T_{ij}^{(n)\omega}(x,y)\,u_j(x)\Big)\,ds(x)$$
$$+ \int_{\Omega}\rho U_{ij}^{\omega}(x,y)f_j(x)\,dv(x) \tag{6}$$

where $I_{\Omega}(y)$ is 1 when $y \in \Omega$ and 0 in other cases.

Numerical solution of equation (6) can be performed by collocation method or by an integral variational approach [8]. When the domain $\Omega$ is infinite and there is no source at infinite distance, it is necessary to give restrictive conditions on the behaviour of the displacement field $u(x,t)$ at infinity. These assumptions are called outgoing Sommerfeld radiation conditions. When there are some sources at infinity (denoted by the field $u^{inc}$), the Sommerfeld conditions are applied to the diffracted displacement field $u^{diff} = u - u^{inc}$.

### 3.4 *Regularization and discretization of the problem*

The integral representation defined by equation (6) is generally not valid for $x \in \partial\Omega$. The formulation of the boundary integral equation along $\partial\Omega$ is then not very easy to obtain as the Green kernels have singular values when $x \in \partial\Omega$. It is then necessary to regularize expression (6) to write the boundary integral equation [2,8].

The problems presented in this article are analyzed in two dimensions (plane or antiplane strains). Two dimensional Green kernels of the infinite space are written using Hankel's functions [8,13]. The regularized solution of equation (6) is estimated by classical boundary finite elements discretization and then by collocation method, that is application of the integral equation at each node of the mesh.

## 4   MODEL FOR A PLANE SH-WAVE

### 4.1 *Modelling of the geological contour*

The East-West geological section depicted in figure 1 is now considered for the numerical simulation. The exact contour of the alluvial basin is described precisely with boundary elements (figure 5). The alluvial layer is supposed to be homogeneous and its mechanical characteristics (and those of the bedrock) are identical to the analytical case (see §2). The excitation is a plane SH-wave with vertical incidence ($\alpha$=90°, first) and various incidences (afterwards). For a plane SH-wave, the motion is anti-plane with respect to the direction orthogonal to the model plane (only one DOF in each node of the mesh).

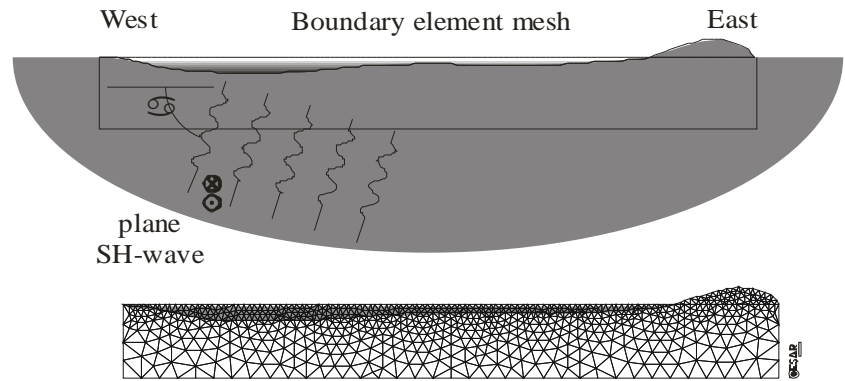

**Figure 5.** Boundary element mesh for a plane SH wave
and points for a posteriori estimation of the solution.

In figure 5, the boundary element mesh is depicted (top) for the solution of the integral equation and a complete mesh (created with a finite element mesh generator, bottom) gives some additional points for the computation of the a posteriori solution inside the domain. This model uses Green's functions of an infinite domain for the alluvial layer and the mount. It involves Green's functions of a semi-infinite domain (easy to determine for SH-waves) to precisely model the bedrock as a subdomain of an infinite half-space [8,13]. The numerical solution is estimated in the frequency domain with the finite/boundary element software CESAR-LCPC [23].

### 4.2 *Amplification factor for a vertical incidence*

Considering the two-layered model, the boundary element method allows for the determination of displacement (and amplification factor) in all points of the boundary element mesh. The solution is determined afterwards in all points inside the domain (figure 5).

Figure 6 gives the isovalues of the amplification factor in the alluvial layer and the bedrock for various frequency values (for a plane SH-wave with vertical incidence). Amplification obviously occurs at the surface of the basin and reaches a maximum value of 15.0 for a frequency value of 1.6 Hz. The maximum amplification appears in the thickest part of the deposit. However, for higher frequencies (f=2.0 Hz and 2.4 Hz), the amplification factor in the thinnest part of the alluvial layer increases slightly or strongly (respectively). For the highest frequencies, results of figure 6 clearly show that maximum amplification occurs in several small areas. This is due to the shorter wavelengthes involved at these frequencies than at lower ones. Maximum amplification areas have therefore a quite limited extent and their experimental determination could be difficult.

 

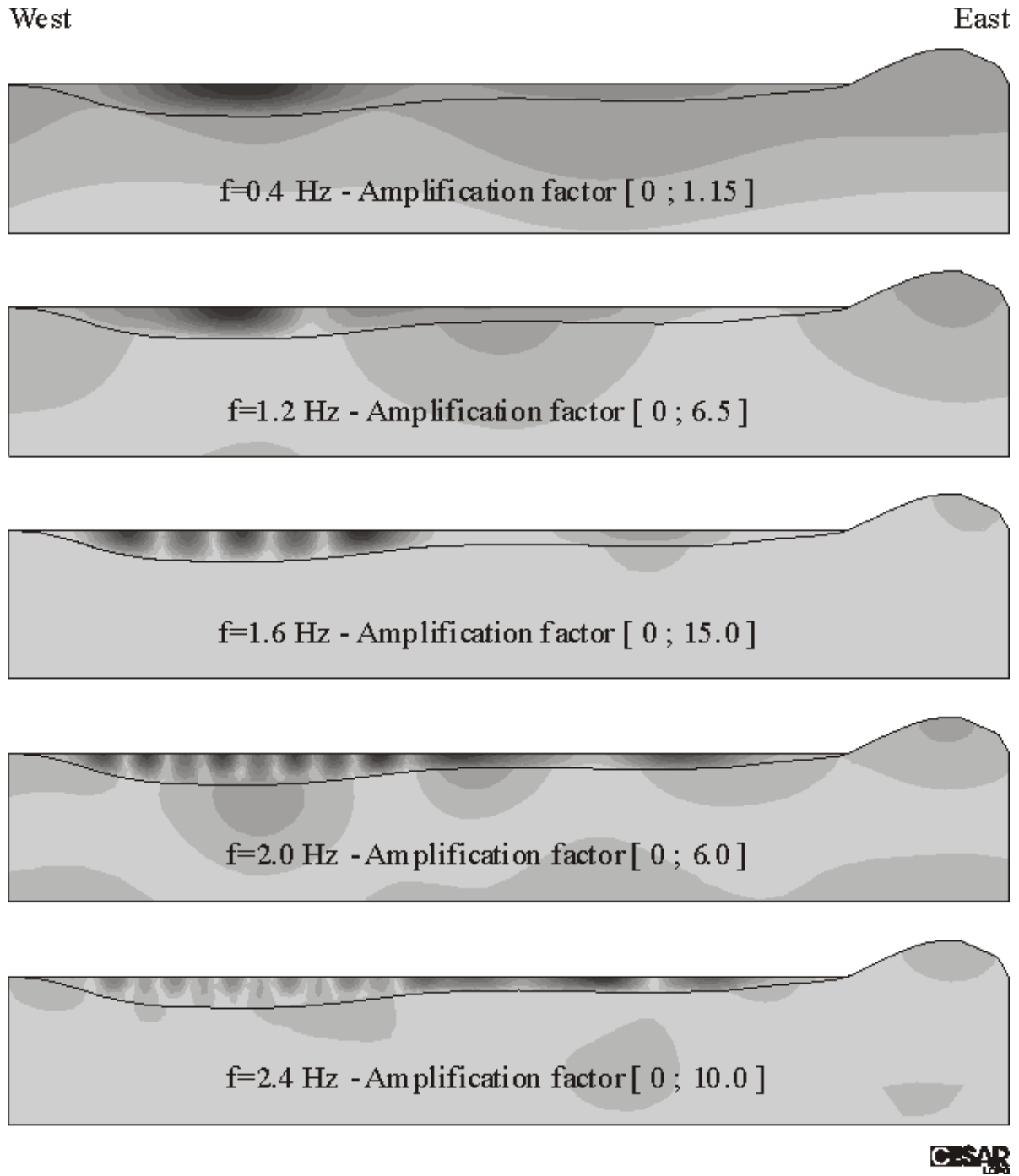

**Figure 6.** Computation of the amplification factor for a vertical SH-wave at different frequencies (variable scale, maximum in black).

Three-dimensional graph of figure 7, and corresponding isovalues plot, give the variations of the amplification factor on the surface of the alluvial basin versus location and frequency. It is then possible to estimate the amplification level at each frequency as well as the location and extent of the corresponding maximum amplification area. In figure 7, the amplification factor is low for frequencies under 0.8 Hz. Above this frequency value, an area of important amplification appears in the thickest part of the alluvial deposit (West). For higher frequencies, several areas of high amplification factor are detected always in the western part of the deposit. Above 1.5 Hz, the amplification factor in the thinnest part of the alluvial basin (East) increases progressively. Between 2.0 and 2.5 Hz, the amplification decreases in the thickest part of the deposit whereas it strongly increases in the thinnest part. As indicated by right part of figure 7, the amplification factor reaches 15.0 or more in the thickest part of the valley for frequencies 1.4 and 1.6 Hz and is between 11.0 and 15.0 in the thinnest part for 2.4 Hz.

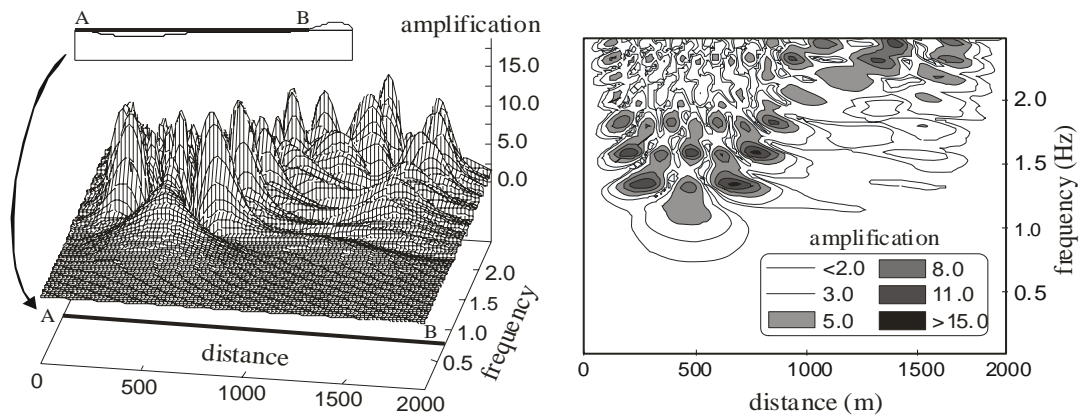

**Figure 7.** Amplification on the surface versus distance and frequency.

Results of figures 6 et 7 indicate that the amplification of the seismic motion reaches a maximum value of 15 for frequencies between 1.4 and 1.6 Hz. The highest amplification firstly occurs in the thickest part of the deposit for moderate frequencies, and then in the thinnest part around 2.4 Hz. These results are in good agreement with the frequency values determined analytically in section 2 (figure 4). Considering an horizontal infinite surface layer appears to be a good assumption for the analytical estimation of the resonant frequencies of the alluvial filling. It is probably because the real alluvial deposit is rather flat with two parts of nearly constant depths (western part : 64 m, eastern part : 32 m).

The amplification level computed with the boundary element model is much higher than that estimated analytically. The numerical estimation of the amplification factor is three times higher than in the analytical case. The numerical model allows an accurate description of the geometry of the deposit leading to a real imprisoning of seismic waves and consequently to stronger site effects. For the fundamental mode, this result is in good agreement with experimental seismic measurements displayed in figure 2. Complete comparisons between experimental and numerical results (in the three directions of space) are proposed in the last section of the article.

### 4.3  *Overall maximum amplification*

In figure 8, the curve displays the overall maximum amplification, that is the maximum amplification factor at each frequency. The location of the maximum amplification is not considered in this curve but is still changing from one frequency to another. As shown in next section, it is not possible to detect all maximum amplification peaks considering the results in only one point of the deposit versus frequency. From the 3D plot of figure 7, we build the curve of overall maximum amplification versus frequency (given in figure 8). The analysis of overall maximum amplification values is necessary to investigate the whole site effects versus frequency.

From the curve given in figure 8, there is no site effects below 0.5 Hz. Between 0.5 and 1.0 Hz, the amplification of seismic motion is limited (less than 4.0). Above 1.0 Hz, the amplification factor increases fastly to reach its maximum value at 1.4 and 1.6 Hz. Site effects are not very strong around 2.0 Hz as the amplification factor is below 7.0. Another strong amplification (around 13.0) is found at 2.4  Hz and was previously shown to appear in the thinnest part of the basin. Weaker site effects also occur around 3.5 Hz but the overall maximum amplification factor decreases progressively with frequency. This global result is compared with experimental ones in the last of the paper as well as for other wave types.

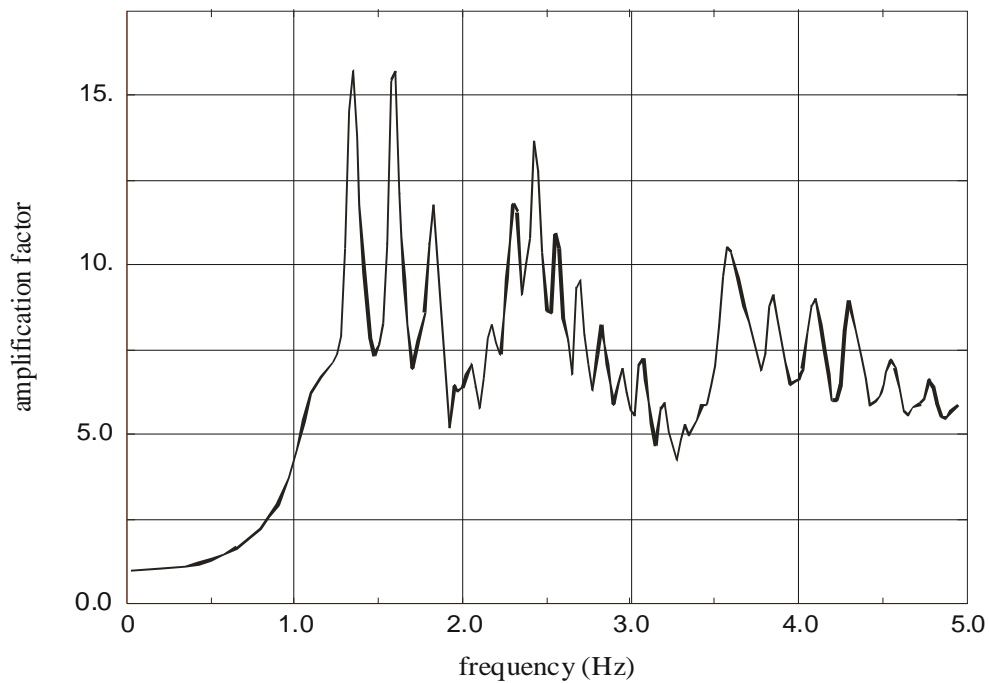

**Figure 8.** Overall maximum amplification factor versus frequency
for a plane SH-wave (variable location).

## 5 INFLUENCE OF INCIDENCE

As the bedrock is described using Green's functions of an infinite half-space, seismic amplification can be determined, in the case of plane SH-waves, for every values of incidence (no truncation effect). To analyze the effect of incidence, results are firstly given in a specific point versus frequency. Afterwards, the amplification factor is studied all along the surface for several values of frequency.

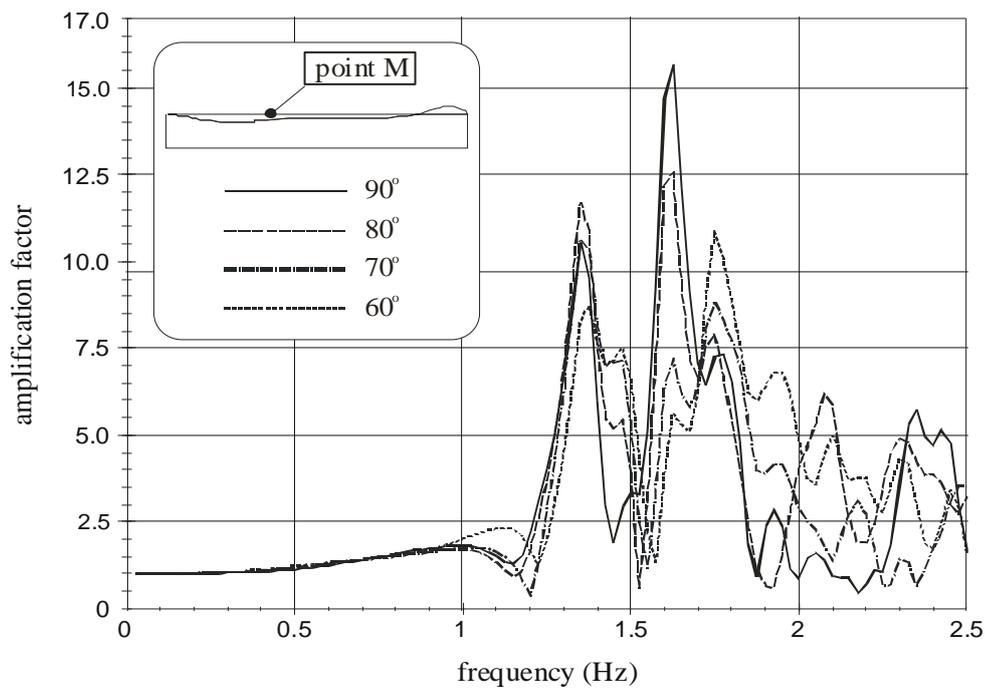

**Figure 9.** Amplification factor versus frequency for different incidences at the point *M* of maximum amplification (plane SH-wave).

Results of figure 9 correspond to the point *M* for which amplification is maximum at frequency 1.6 Hz (see figures 6 et 9). This point *M* is located at distance $d_M$=860 m of the western edge of the basin. From these curves, amplification is obviously very low for frequencies under 1.2 Hz. Above this frequency value, amplification factor strongly increases (between 10 and 12 depending on incidence). Around 1.5 Hz, amplification decreases and increases once more up to its maximum value for an approximate frequency of 1.6 Hz. For this specific frequency, amplification factor decreases when incidence angle is low. Point *M* corresponds to the maximum amplification for a normal incidence but is probably not the point of largest amplification for other incidences. When comparing figure 9 to figure 8, it can be noticed that seismic site effects are strongly underestimated in figure 9 for frequencies below or above 1.6 Hz. That is why the estimation of the overall maximum amplification is needed. The location of the maximum amplification factor is different from one frequency to another. The following results will allow comparison of the locations of site effects for various incidences.

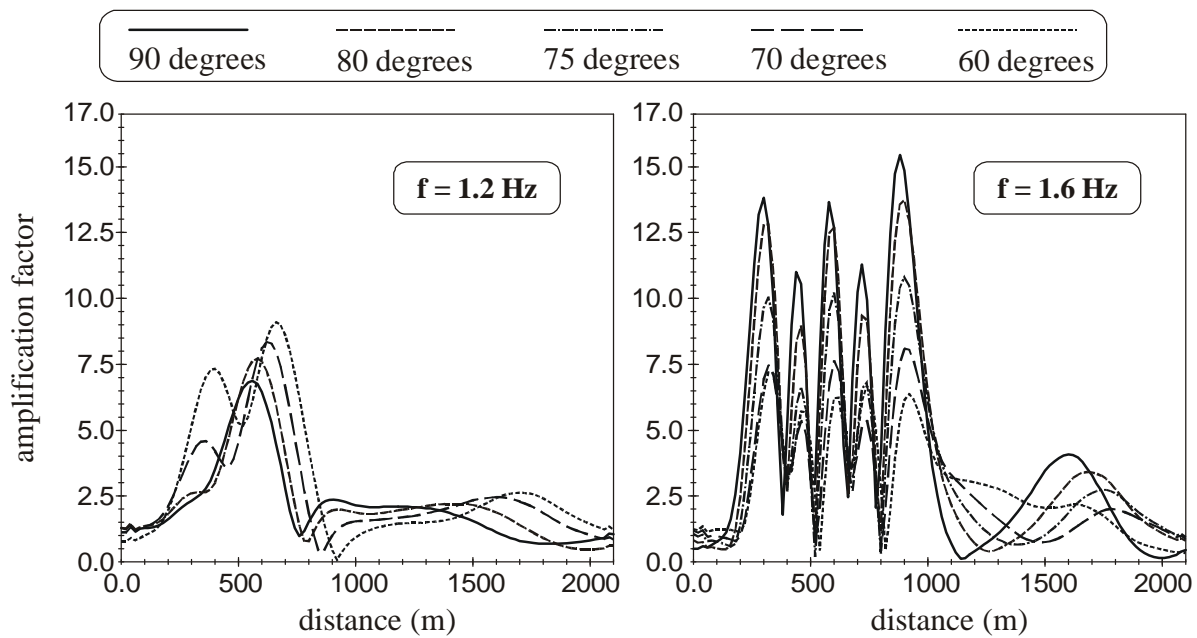

**Figure 10.** Dependency of amplification factor on incidence angle.

Results of figures 10 give the values of the amplification factor in all points of the basin surface for two specific frequency values. It is then possible to investigate the influence of incidence on site effects and their location. From figure 10, it can be seen that the amplification factor at 1.2 Hz increases with decreasing incidence : it has a value of 7.0 for a normal incidence (90°) and reaches 9.0 for a 60° incidence. The location of the maximum amplification area has significant changes : it is at a distance of 500 m from the western edge of the deposit for a normal incidence and at 700 m for a 60° incidence (figure 10). The numerical modelling gives an interesting estimation of spatial variability of site effects [12,33].

For a frequency of 1.6 Hz (figure 10), the amplification factor decreases when incidence is lower. Its value is 15.5 for a normal incidence but is under 7.0 for a 60° incidence. It is also interesting to notice that, for a 60° incidence, seismic site effects are stronger at 1.2 Hz than at 1.6 Hz. The resonant frequency of the surface layer is still changing from one incidence to another. The differences between results corresponding to various incidences are due to the dependence of horizontal and vertical wavenumbers on incidence. The deposit « appears » thicker for decreasing incidence and the amplification factor for low frequencies is then higher for lower incidences. Furthermore, both basin edge and focusing effects [42] are certainly affected by the incidence angle and then modify the spatial variability of site effects.

# 6  AMPLIFICATION IN A DAMPED MEDIUM

## 6.1  *Damping properties of the model*

The boundary element model presented previously does not involve any damping. In this section, the numerical model is supposed to have a linear viscoelastic behaviour. The formulation of damping corresponds to a Zener model or standard solid [11,35,36]. This rheological model is depicted in figure 11 for shear response and the expression of the inverse of the quality factor $Q^{-1}$ (i.e. attenuation) is given as a function of frequency. The variations of $Q^{-1}$ with frequency are also drawn in this figure showing a peak corresponding to the maximum value of attenuation. Considering this rheological model, the complex shear modulus of the medium $\mu^* = \mu_R + i.\mu_I$ can be written as a function of frequency, short term (instantaneous) $\mu_{st}$ and long term $\mu_{lt}$ shear moduli as follows :

$$\begin{cases} \mu_R = \dfrac{\mu_{lt}\mu_{st}^4 + \omega^2\zeta^2\mu_{st}\left(\mu_{st} - \mu_{lt}\right)^2}{\mu_{st}^4 + \omega^2\zeta^2\left(\mu_{st} - \mu_{lt}\right)^2} \\[3mm] \mu_I = \dfrac{\omega\zeta\mu_{st}^2\left(\mu_{st} - \mu_{lt}\right)^2}{\mu_{st}^4 + \omega^2\zeta^2\left(\mu_{st} - \mu_{lt}\right)^2} \end{cases} \tag{7}$$

where $\zeta$ is the viscosity coefficient of the Zener's model (figure 11), $\mu_R$ and $\mu_I$ are the real and imaginary parts of the complex shear modulus and the long term shear modulus is such as :

$$\frac{1}{\mu_{lt}} = \frac{1}{\mu_{st}} + \frac{1}{\mu'}$$

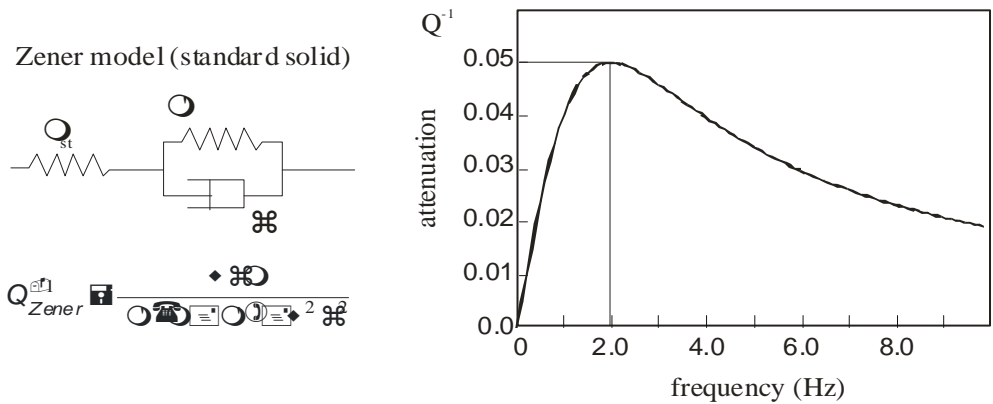

**Figure 11.** Zener's model (standard solid) and corresponding attenuation vs frequency dependency.

For the computation of wave amplification in the damped surface layer, the characteristics of Zener model are chosen such as the maximum attenuation value is reached at a frequency of 2.0 Hz. The elastic bedrock is supposed to be undamped. The values of attenuation $Q^{-1}$ and of corresponding damping ratio $\xi$ ($Q^{-1}=2\xi$) for the basin are given in table I for the three cases studied.

**Table I** : Maximum values of the inverse of the quality factor and the damping ratio.

|  | undamped | case 1 | case 2 | case 3 |
|---|---|---|---|---|
| max. $Q^{-1}$(at 2 Hz) | 0.00 | 0.02 | 0.04 | 0.06 |
| damping ratio $\xi$ |  | 1 % | 2 % | 3 % |

## 6.2 *Amplification and damping*

The amplification factor along the basin surface is firstly studied. Figure 12 gives its values at several frequencies and for different attenuations. For the four frequencies considered, the values of attenuation $Q^{-1}$ can be computed from the expression given in figure 11. For 2.0 Hz, attenuation values are already given in table I. For the other frequencies, its values are the following for the case of highest damping (No.3) : $Q^{-1}$(1.0 Hz)=0.048, $Q^{-1}$(1.4 Hz)=0.056 and $Q^{-1}$(1.6 Hz)=0.059. For 1.0 Hz, the amplification level is rather low and the influence of damping is not very strong. At this frequency, there is no effect of damping on the location of maximum amplification. For 1.4 and 1.6 Hz, site effects are much stronger and the influence of damping on amplification factor values can be large since, for the highest attenuation value, they are divided by a factor of 2 at 1.4 Hz and by a factor of 3 at 1.6 Hz.

However, for the lowest attenuation value, the discrepancy between damped and undamped curves is not very large. Near the so-called resonance of the basin, the influence of damping on amplification can be very strong for high attenuation values. For both frequencies, there is still no significant effect on the location of the peaks. For 2.0 Hz, that is the frequency giving the maximum value of attenuation ($Q^{-1}$=0.06), the influence of damping on amplification is lower than for 1.4 and 1.6 Hz. The location of maximum amplifications is however significantly modified between the undamped case and the case of highest damping.

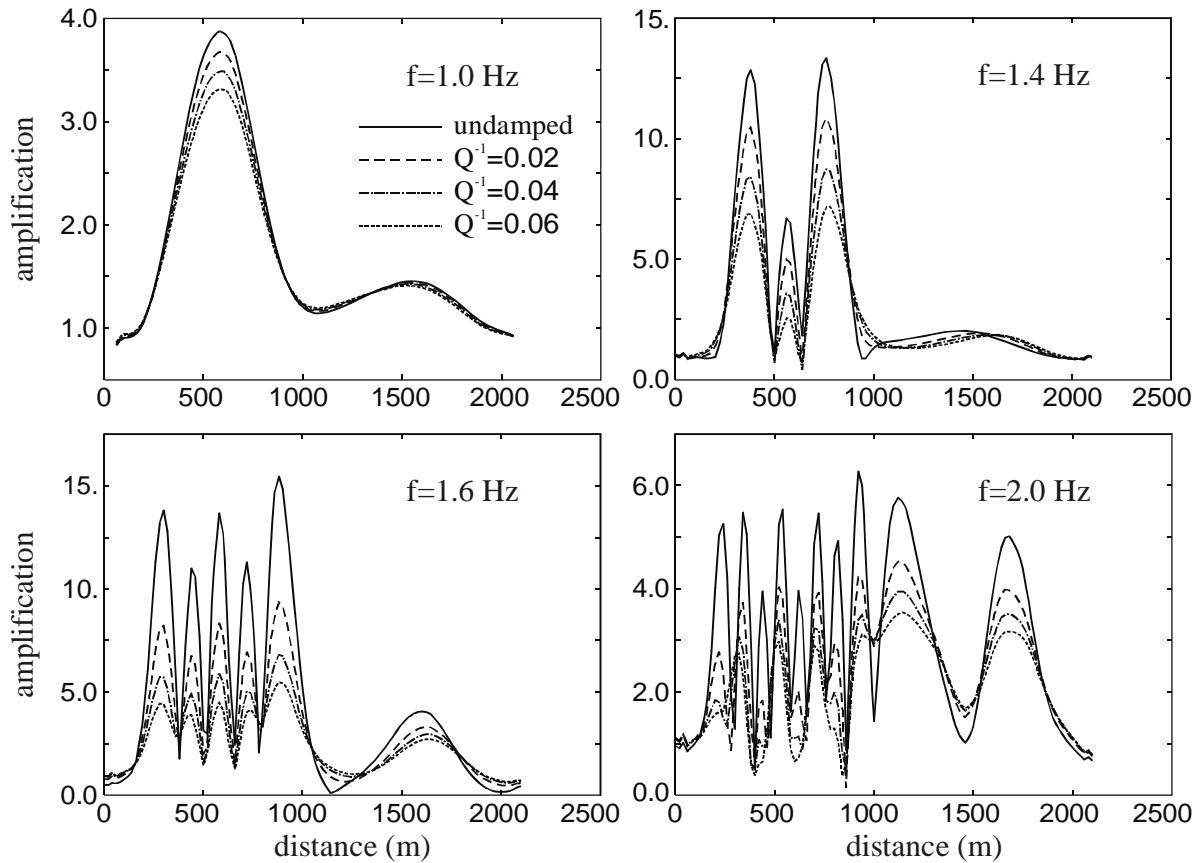

**Figure 12.** Amplification factor along the basin surface
in damped and undamped cases for different frequencies.

It is interesting to investigate the variations of the overall maximum amplification vs frequency for different values of damping. Figure 13 gives the maximum amplification factor for the undamped case and for the three damped cases presented in table I. From these curves, the overall maximum amplification factor appears to have significant differences between damped and undamped cases. The relative decrease of the amplification factor in the three damped cases (No.1, 2 and 3) compared with undamped case is respectively -28 %, -45 % and -54 %. For the highest attenuation value ($Q^{-1}$=0.06), the difference is very large between damped and undamped cases but it appears smaller than in figure 12 since there is a significant shift in the amplification peaks. As far as the amplification level is concerned, the comparisons made in figure 12 are not completely satisfactory since the amplification peaks are not related to the same frequencies. Concerning the frequency dependency of the amplification factor, similar amplification types are observed in the undamped cases, that is three main amplifications around 1.4/1.6 Hz, 2.4 Hz and between 3.5 and 3.8 Hz. The frequency shift of the amplification peaks in the damped cases is much larger for the third main amplification. The curves of figure 12 and 13 show the strong effect of damping on amplification level but a very slight influence on occuring frequencies. It is therefore very important to perform, in the specific site considered, reliable measures of the damping properties of the surface soil layers.

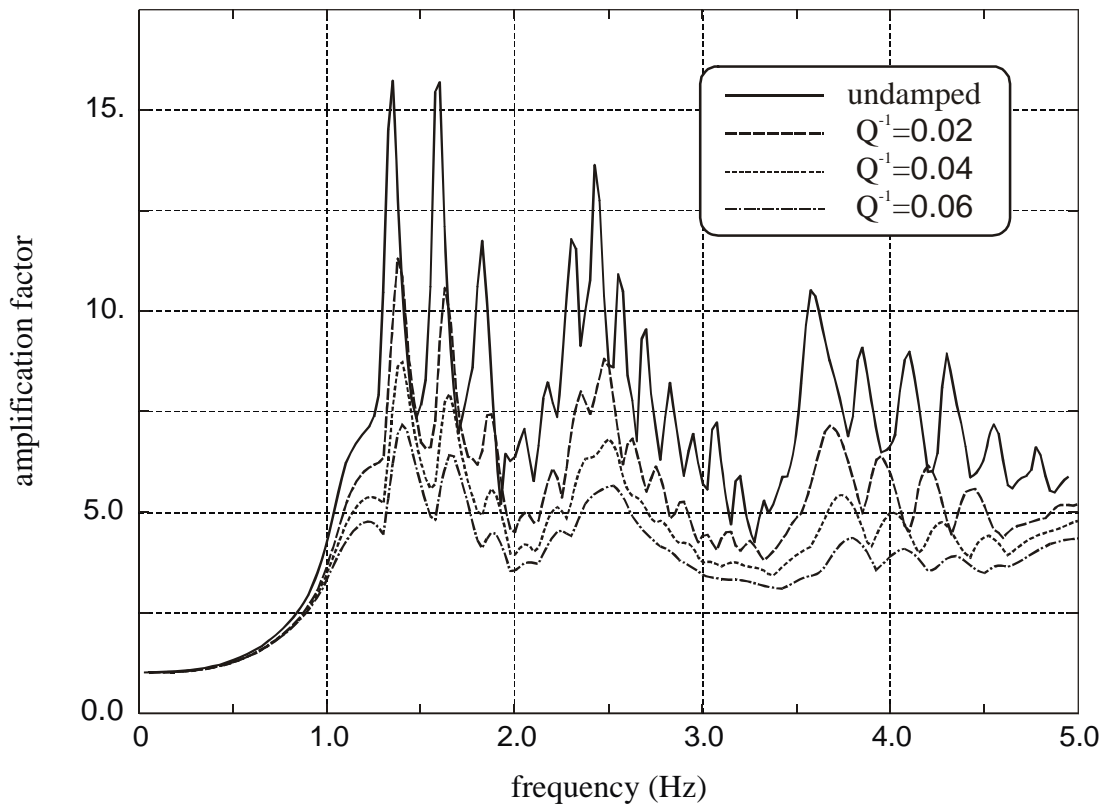

**Figure 13.** Overall maximum amplification factor vs frequency for a damped and undamped surface layer.

## 7 P AND SV-WAVES

For pressure and shear waves (P and SV), the motion is in the propagation plane. It is not possible to use Green's functions of an infinite half-space (as for SH-waves) and one needs to extend the boundary element mesh on both sides of the deposit. The boundary element mesh considered in the case of P and SV waves is given in figure 14. To avoid numerical errors due to the truncation of the mesh, it has to be sufficiently extended to model correctly the infinite wideness of the bedrock. P and SV waves computations are performed using several punctual sources at a 500 m depth. The amplitude of these sources is arbitrary. The amplification factor is estimated in reference to preliminary computations only involving an homogeneous half-space.

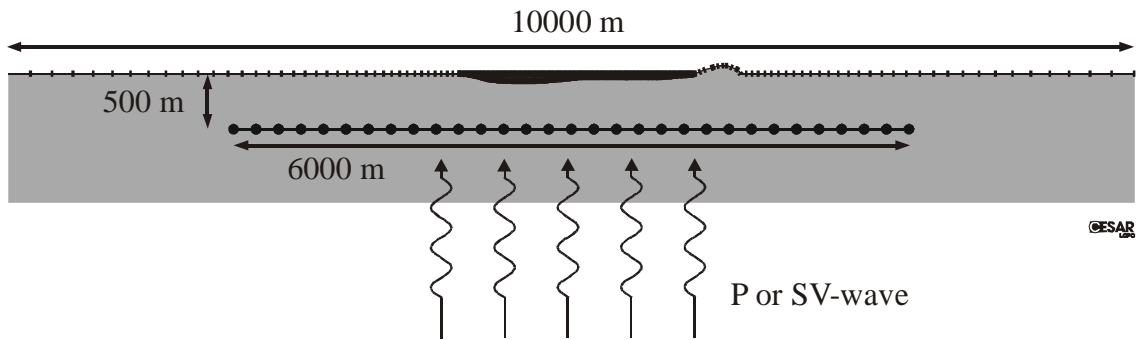

**Figure 14.** Boundary element mesh for P and SV waves.

Curves of figure 15 give the amplification factor values in every point of the surface for various frequencies and for three different wave types (SH, P and SV). For these simulations, mechanical characteristics of the deposit and of the bedrock are the same as in sections 2 and 4.

For low frequencies (1.2 and 1.4 Hz, figure 15), P and SV-wave amplification is very low whereas SH-waves amplification is significant (between 7.0 and 8.0), and even stronger at 1.4 Hz (13.0). The maximum amplification always occurs in the thickest part of the alluvial deposit.

For higher frequencies (1.6 and 1.8 Hz, figure 15), seismic amplification of SH-waves is maximum (15.5) for 1.6 Hz whereas it is below 4.0 for P and SV waves. At 1.8 Hz, site effects are rather equivalent for the three wave types and the amplification factor is between 6.0 and 8.0. The highest amplification factor is always reached in the western (thickest) part of the deposit. Nevertheless, the number and extent of maximum amplification areas depend on the wave type. Around 2.0 Hz, P- and SV-waves give strong amplification (13.0 and 14.0 respectively) but for SH-waves, the amplification factor is under 6.0. At 2.2 Hz, the amplification in the thinnest part of the deposit increases for SH-waves (7.5). In the thickest part of the deposit, the amplification factor is always under 8.0 for all wave types.

For all frequencies, the maximum amplification factor is then between 13.0 for P-waves and 14.0 for SV-waves. The highest amplification factor is reached for SH-waves at 1.6 Hz (16.0). It is always reached in the thickest part of the deposit, whereas the thinnest part has lower amplification at higher frequencies.

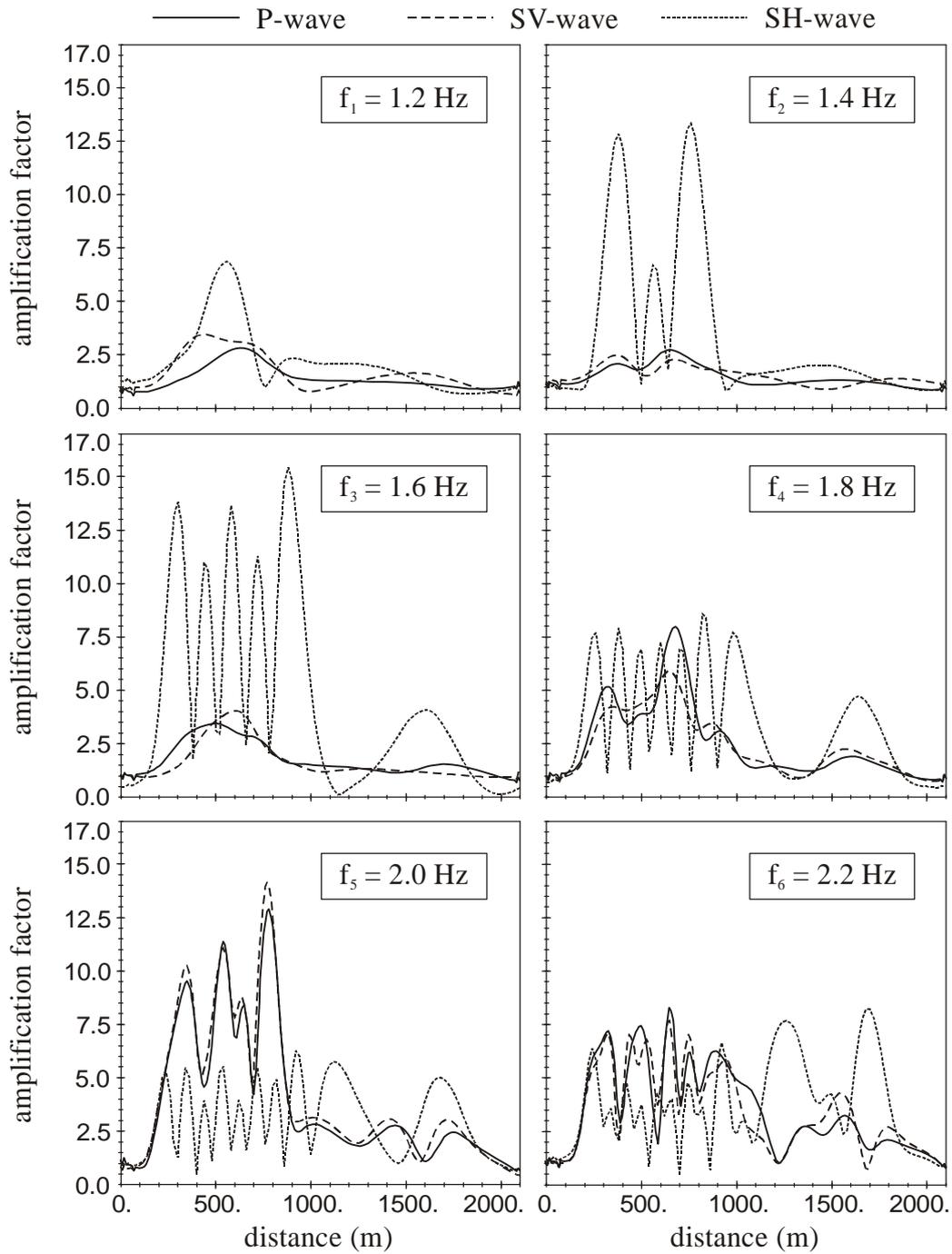

**Figure 15.** Amplification factor for different wave types at various frequencies.

## 8 COMPARISON BETWEEN NUMERICAL AND EXPERIMENTAL RESULTS

### 8.1 *Mechanical parameters of the alluvial filling*

As the alluvial deposit is assumed to be an homogeneous layer of constant shear modulus, it is easy to analyze the influence of the mechanical characteristics of the basin on the amplification process. In the next section, this preliminary analysis will enable a complete comparison between numerical and experimental results. Three different values of the shear modulus of the basin are considered in this section : $\mu_1$=180 MPa as for the previous computations, $\mu_2$=2$\mu_1$/3=120 MPa and $\mu_3$=$\mu_1$/2=90 MPa. The corresponding shear wave velocities are the following : $C_1$=300 m.s$^{-1}$, $C_2$=245 m.s$^{-1}$ and $C_3$=212 m.s$^{-1}$.

The curves given in figure 16 show the amplification levels along the basin surface for these values of shear modulus at two different frequencies. At 1.2 Hz, the number of amplification peaks is larger for lower shear moduli since the corresponding wave velocities are lower and lead to smaller wavelengthes. Values of amplification factor are different in the three cases and it ranges from 7.0 in the first case ($\mu_1$) to 10.0 in the third case ($\mu_3$). Furthermore, the maximum values are not reached at the same places along the basin surface. At 1.8 Hz, the discrepancy between the three curves is even larger since the maximum amplification ranges from 8.5 in the first case to 15.0 in the third case. The location of maximum amplification areas is also very different : in the first case, amplification is maximum in the western part, in the second case, amplification is maximum in the middle of the basin and in the third case, amplification is maximum in the eastern part. These results show the importance of in situ measurements to estimate geometrical and mechanical characteristics of the surface layers. These characteristics are mean parameters estimated experimentally. The numerical model allows the analysis of sensitivity to these parameters. These different numerical results are compared hereafter with experimental ones.

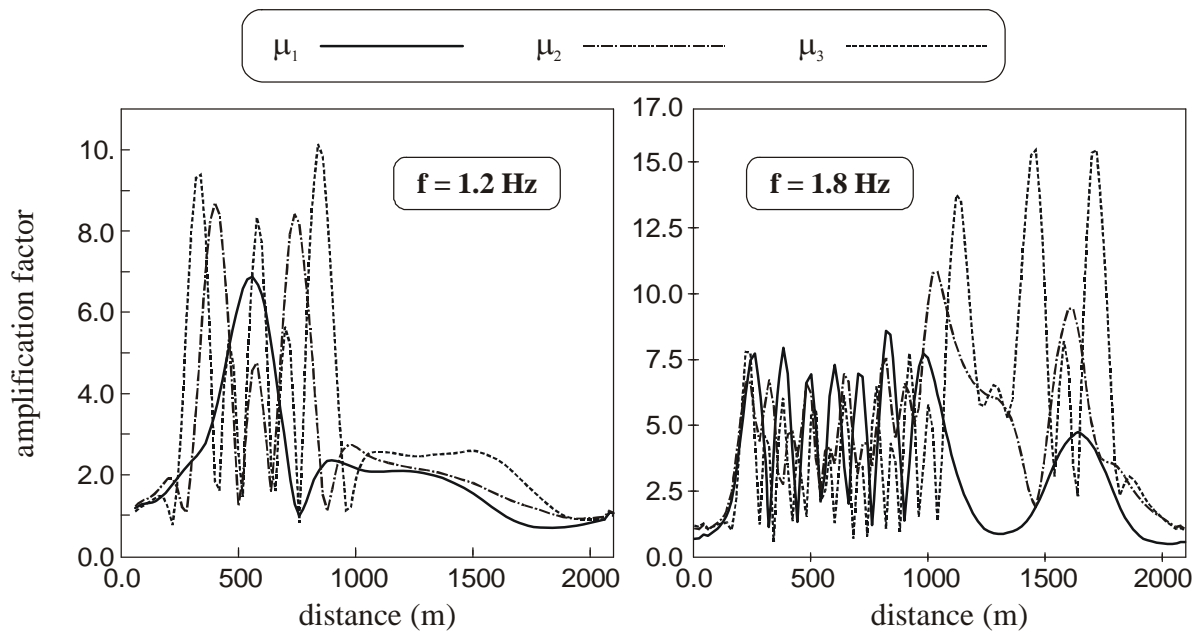

**Figure 16.** Amplification factor versus distance for three values
of shear modulus of the alluvial basin at two different frequencies.

8.2  *Model/experiments comparisons in the three directions of space*

The numerical results are now compared with experimental measurements of seismic
site effects in the case of real earthquakes (weak motion). We consider the experimental re-
sults displayed in figure 2 giving the spectral ratios (site/reference) for the three motion
components : vertical, North-South, East-West. Spectral ratios of the three different com-
ponents of the seismic motion are respectively compared to numerical results with P-wave,
SH-wave and SV-wave loading (with normal incidence). Since the geological profile is
along East-West direction (figure 1), vertically incident P-wave gives vertical motion, SH-
wave corresponds to anti-plane motion in North-South direction and vertically incident SV-
wave is related to East-West motion.

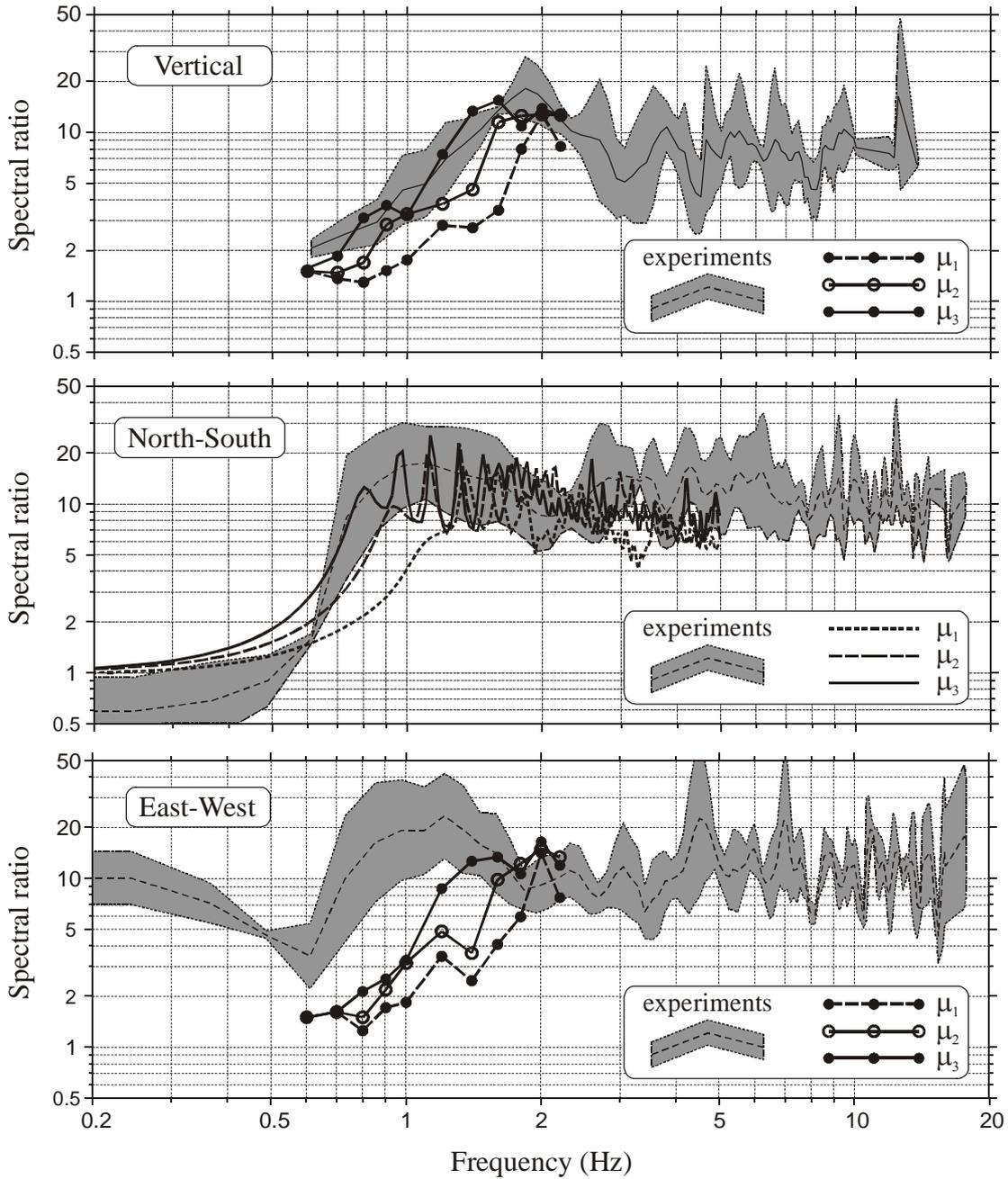

**Figure 17.** Comparisons between experimental measurements (real earthquakes) and numerical results in the three motion components.

Figure 17 gives both numerical and experimental results : numerical results correspond to the overall maximum amplification and, for experimental ones, the dotted line gives the mean value and the shaded area the standard deviation interval. The comparisons for each motion component lead to the following conclusions :

- *vertical component* : the numerical results are computed for eleven frequency values between 0.6 and 2.2 Hz (the number of frequencies is limited since one has to estimate both amplitudes in the case involving the alluvial basin and in the case of an infinite half-space). For the vertical motion component, that is P-wave excitation for boundary element model, the three numerical results are not far from the experimental curves. The third case (lowest shear modulus) leads to a very good agreement with the experimental spectral ratio since all ratios estimated numerically are within the standard deviation range of the experiments (at the same frequencies),

- *North-South component* : the numerical results are computed for 200 different frequency values. For the North-South motion component, that is SH-wave excitation for boundary element model, the three numerical results are rather close to the experimental curves. The third case ($\mu_3$) gives very good numerical results concerning maximum amplification levels when compared to the experimental spectral ratios. The two main amplification in the thickest part of the deposit (above 1.0 Hz) and in the thinnest part (above 2.5 Hz) are detected with both approaches. These frequencies are not far from those determined analytically (§2) assuming an infinite alluvial layer of constant depth. Nevertheless, for all three numerical cases, the experimental spectral ratios are smaller than numerical amplification factors for low frequencies (below 0.6 Hz),

- *East-West component* : the numerical results are computed for eleven frequency values between 0.6 and 2.2 Hz. For the East-West motion component, that is SV-wave excitation for boundary element model, the three numerical results are not very close to the experimental curves. The third case ($\mu_3$) leads to the best agreement with the experimental spectral ratios but the discrepancy is rather significant since it is about 0.4 Hz in terms of frequency shift. Nevertheless, for all three numerical cases, the numerical amplification factors are of the same order than the experimental spectral ratios but for higher frequencies.

For vertical and North-South motion components, numerical results are in very good agreement with experimental ones giving close amplification levels at same frequencies. For both components, the best agreement is obtained in the third case, that is the lowest shear modulus ($\mu_3=\mu_1/2=90$ MPa) or wave velocity ($C_3=212$ m.s$^{-1}$). In the case of SV-waves (East-West component), the maximum amplification factor estimated numerically is as high as the experimental spectral ratio but there is a frequency shift of approximately 0.4 Hz. For this motion component, the assumption of a homogeneous alluvial basin is perhaps not appropriate and one has probably to precisely model the velocity distribution with depth.

## 9  CONCLUSIONS

Site effects lead to a local amplification of seismic motion. They are significant in the highly builded areas of the center of Nice. In situ experiments performed by the CETE-Méditerranée show that the seismic motion should be strongly amplified between 1 and 2 Hz. The amplification levels are around 20 and cannot be recovered considering a simple analytical model involving an infinite surface layer of constant depth.

Numerical simulations based on the boundary element method allows a precise description of the site as well as an accurate analysis of seismic wave propagation within the alluvial basin. Amplification levels and occuring frequencies are of the same order as the experimental values. As also shown by the experiments, the maximum amplification estimated numerically is located in the thickest part of the alluvial basin for lower frequencies and in the thinnest part for higher frequencies.

Site effects quantified numerically are sensitive to incidence. It changes the maximum amplification factor, the frequency at which it occurs and the corresponding location. The influence of damping on amplification is also large. Two types of interpretation of this strong influence could be proposed : the first one in terms of vibration and resonance considering that the influence of damping near the resonance of a system is large, the second one in terms of wave propagation taking into account the multiple wave reflections in the alluvial layer increasing consequently the effect of damping.

Finally, for the various wave types considered (SH, P, SV), seismic motion amplification is very different but always reaches a high level (between 13.0 and 16.0). Shear waves lead to the strongest site effects. The location of maximum amplification areas is also depending on the features of the seismic loading. The mechanical characteristics of the deposit have a significant influence on these numerical results. For SH-waves, the amplification factor ranges from 16.0 at 1.35 and 1.60 Hz (for the highest shear modulus) to 25.5 at 1.12 Hz (for the lowest modulus). The lowest shear modulus leads to numerical results in very good agreement with experimental ones for P- and SH-waves (vertical and North-South components respectively).The boundary element method seems to be efficient to analyze site effects from a qualitative as well as quantitative points of view.

To improve the numerical boundary element model, two main points would have to be emphasized : the description of the different surface layers with variable shear modulus and the influence of their respective damping features. These two points may lead to a better understanding of seismic site effects and allow more detailed comparisons between experimental and numerical results. Nevertheless, the experimental determination of *dynamic* mechanical properties is often difficult (costly, many different methods (cyclic, dynamic...), scattered values) especially when analyzing their distribution with both depth and distance along a complete geological profile.

## 10 ACKNOWLEDGEMENTS

The authors are indebted to Dr Pierre-Yves Bard (LCPC / LGIT Grenoble University) for his pertinent remarks on the preliminary version of this manuscript.